# Calculation of spatial response of 2D BES diagnostic on MAST[a]

Young-chul Ghim(Kim),[1,2] A. R. Field,[2,b] S. Zoletnik[3] and D. Dunai[3]

[1]*Rudolf Peierls Centre for Theoretical Physics, University of Oxford, Oxford, UK*
[2]*EURATOM/CCFE Fusion Association, Culham Science Centre, Abingdon, UK*
[3]*KFKI RMKI, Association EURATOM/HAS, Budapest, Hungary*

The Beam Emission Spectroscopy (BES) turbulence diagnostic on MAST is to be upgraded in June 2010 from a 1D trial system to a 2D imaging system (8 radial × 4 poloidal channels) based on a newly developed APD array camera. The spatial resolution of the new system is calculated in terms of the point spread function (PSF) to account for the effects of field-line curvature, observation geometry, the finite lifetime of the excited state of the beam atoms, and beam attenuation and divergence. It is found that the radial spatial resolution is ~ 2-3 cm and the poloidal spatial resolution ~ 1-5 cm depending on the radial viewing location. The absolute number of detected photons is also calculated, hence the photon noise level can be determined.

## I. INTRODUCTION

Beam Emission Spectroscopy (BES) can be used as a technique for measurement of density fluctuations up to a few 100 kHz in magnetically confined plasmas.[1] Durst et al. measured the characteristics of density fluctuations in TFTR such as the radial profile, the poloidal group velocity, the correlation length and the decorrelation time.[2] These results have provided valuable data on the characteristics of tokamak turbulence which is believed to cause anomalous transport. McKee et al. installed a 2D BES system on DIII-D and reported density fluctuation measurements at a poloidal cross-section.[3] Since then, there have been numerous interesting results on DIII-D from the 2D BES, for example: detection of geodesic acoustic modes (GAM),[4] observation of zero-mean-frequency (stationary) zonal flows,[5] and density fluctuation suppression during the formation of internal transport barriers (ITB).[6] It is recognized that the confinement of the plasma improves as the ratio of zonal flow power to drift wave power increases.[7] Also, Shi has predicted theoretically that at low aspect ratio the level of zero-mean-frequency zonal flow should be higher.[8]

With the aim of performing such investigations on MAST, a 1D (8 radial channels) BES trial system has been installed on the Mega-Amp Spherical Tokamak (MAST). This has yielded data on the turbulence characteristics of L-mode plasmas, coherent MHD in H-mode plasmas and precursor oscillations to edge-localized modes (ELMs).[9] In June 2010, the BES on MAST is to be upgraded from the 1D trial system to 2D imaging system (8 radial × 4 poloidal channels) based on an APD array camera. The 1D trial system is limited to detect fluctuations of a few 1%, but the new 2D system has much higher étendue (a factor of 35 larger) and will be able to observe fluctuations of a few 0.1 % at 1 MHz bandwidth. Quantitative interpretation of BES data in space and time requires a forward model of the 2D BES system. Such a model provides calculations of the spatial response in the form of point spread functions (PSF) and the level of photon noise, which is expected to dominate the signal-to-noise ratio. This information can be used to implement a synthetic BES diagnostic operating on fluctuation data from numerical turbulence simulation codes the output of which can then be directly compared with the measurements.

## II. GENERATING 2D SYNTHETIC BES DATA

The new 2D BES system on MAST is designed to observe emission from one of the NBI heating beams (SS) with injection energy of 60 – 75 keV. The system detects the $D_\alpha$ emission from the collisionally induced excited beam atoms which is Doppler shifted by ca. 3 nm away from background $D_\alpha$.[9] The viewing location can be radially scanned with a direct coupled imaging system whose focal point follows the axis of the SS beam. The poloidal locations of the views are fixed at Z = -0.03, -0.01, 0.01 and 0.03 m. The effective f-number at the beam is F/10 and F/1.2 at the detector with the étendue (optical throughput) of the collection optics of $\Delta\xi \sim 1.1 \times 10^{-6}$ m$^2$ sr.[9] The APD array camera system has an active area of $1.6 \times 1.6$ mm$^2$. With a magnification factor of 8.7, the area of each channel is $1.5 \times 1.5$ cm$^2$ with 2 cm separation between centres of adjacent channels at the axis of the SS beam at a nominal radial location of 1.2 m. A detailed description of the system can be found elsewhere.[9] Without considering other physical effects, the spatial response will be limited to wave-numbers $k_r$, $k_\theta$ (or $k_z$) < 1.6 cm$^{-1}$ which should be sufficient to detect long wavelength turbulence given the typical ion Larmor radius $\rho_i \sim 1$ cm on MAST. The actual spatial response has to be calculated by considering relevant physical effects such as magnetic field-line curvature, line-of-sight (LoS) geometry, finite half-life of $D_\alpha$, and the beam attenuation and divergence. For this first generation of 2D synthetic BES data, plasma density (average and fluctuating parts) and temperature profiles are provided by the global, fully electromagnetic, two-fluid nonlinear code CUTIE.[10]

BES is a volume-sampling diagnostic, therefore in order to generate synthetic BES data one really requires LoS integration based on 3D input data of the fluctuating plasma density which would require very large data files. However, by availing of the 2D nature of the turbulence the synthetic BES data can be constructed with 2D input data on the poloidal cross-section at a fixed toroidal angle, i.e. the fluctuating plasma densities are highly elongated along the magnetic field-lines with short perpendicular correlation lengths ($k_\perp / k_\parallel \gg 1$).[11] In fact, the BES measurements are spatially localized to a good degree because the LoS at the intersection of the beam are approximately parallel to the local field-lines and the intersection length is much shorter than the parallel correlation length of the turbulence. Nevertheless, as the SS heating beam has 1/e half-width of 8 cm, LoS integration will smear the spatial information to some extent.

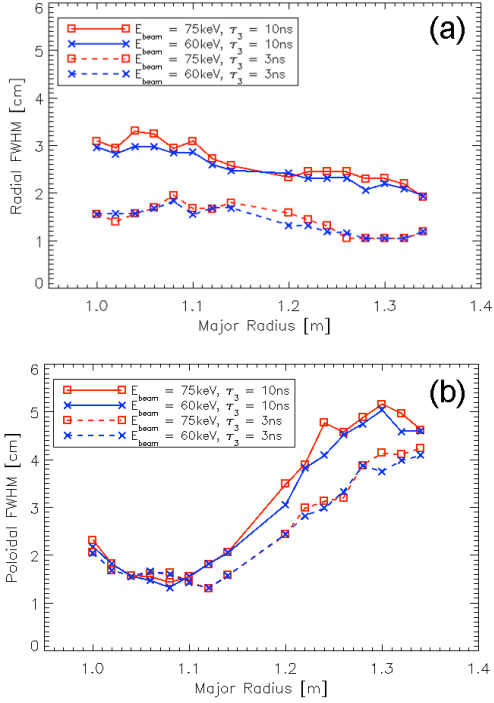

FIG. 1. (Colour online) (a) Radial and (b) Poloidal profiles of the FWHM of the PSF as functions of beam energy and half-life ($\tau_3$) of n=3-2 transition of neutral deuterium.

In order to calculate the spatial smearing from such 2D input data, successive image planes are constructed along the LoS.[12] The size of the detector images at these planes is set by the optical magnification factors which vary along the LoS. Light-cones whose sizes are determined by the optical system are convolved with these detector images. The resulting blurred images are then convolved with an exponentially decaying function whose e-folding distance is calculated by considering the beam velocity projection of the perpendicular to the LoS and the half-life of $D_\alpha$. Hutchinson calculated the half-life of $D_\alpha$ ($\tau_3$) with the collisional transitions as well as the coronal description of the excited state.[13] It is reported that at a plasma density of $\sim 10^{19}$ m$^{-3}$, $\tau_3$ is $\sim$3-10 ns. The resultant images along the LoS are then moved to the optical focal plane by following the field lines displacing the 2D fluctuation data appropriately. Because the signals at each displaced image plane are integrated at the poloidal cross-section where the optical focal plane is located, plasma densities on each plane can be calculated with the 2D input data from CUTIE[10] which have 500 time points with 1 MHz frequency.

The local beam densities are calculated to estimate the photon detection rate using the ADF21 module for the stopping rate and the ADF22 module for the excitation rate from the ADAS atomic database. This procedure is similar to the calculation method in Ref. [9] except that the simulation can now take account of the fact that beams on MAST consist of multiple beamlets. To reduce the computational time, multiple beamlets are not considered to produce the results presented here.

Fig. 1 shows the radial and poloidal FWHM profiles as a function of the primary beam energy ($E_{beam}$) and finite half-life ($\tau_3$) for $R_{major} = 0.99 - 1.15$ m and $R_{major} = 1.19 - 1.35$ m. Spatial smearing is weakly dependent on $E_{beam}$ because the velocity of the beam is proportional to $E_{beam}^{1/2}$. Altering the half-life ($\tau_3$) from

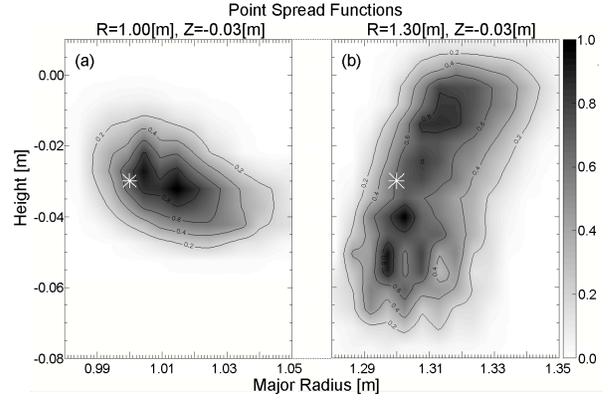

FIG. 2. Point spread functions at two different major radii ($R_{major}$), (a) $R_{major} = 1.00$ m and (b) $R_{major} = 1.30$ m. A white asterisk symbol in each figure shows the focal point of the imaging system.

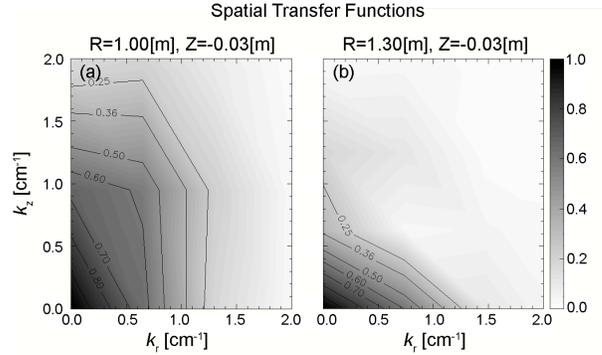

FIG. 3. Spatial transfer functions calculated from the point spread functions shown in Figure 2. Because spatial transfer functions vary strongly with major radius, deconvolution of the measured signal is particularly difficult.

3 to 10 ns changes the radial FWHM by a factor of 1.5 whereas a change of only 1.2 is seen in the poloidal direction at large $R_{major}$ because the beam is moving radially rather than poloidally. Unlike large aspect ratio tokamaks, variation of the magnetic pitch angle is not negligible from the core to the edge on MAST, and this effect can be seen in the strong dependence of the poloidal FWHM of the PSF as a function of viewing radius.

## III. RESULTS AND DISCUSSIONS

Two sets of 2D synthetic BES data with $E_{beam} = 75$ keV and $\tau_3 = 10$ ns provide PSFs for $R_{major} = 0.99 - 1.15$ m ($\psi_N = 0.05 - 0.3$) and $R_{major} = 1.15 - 1.31$ m ($\psi_N = 0.3 - 0.8$) at fixed poloidal locations where $\psi_N$ is the normalized poloidal flux. Note that $\tau_3 = 10$ ns rather than 3 ns is used to obtain the upper limit of the BES spatial resolution on MAST. Fig. 2 depicts PSFs at the simulated minimum ($R_{major} = 1.00$ m) and maximum ($R_{major} = 1.30$ m) major radii at Z = -0.03 m. For ease of comparison, the PSFs are normalized to their own maximum responses. A white asterisk symbol on each figure shows the optical focal point. Due to the effect of finite half-life and the magnetic pitch angle, these focal points do not coincide with the maximum response of the PSFs. As seen in Fig. 1, the PSF at $R_{major} = 1.30$ m shows larger poloidal spatial smearing than at $R_{major} = 1.00$ m. Fig. 3 shows the spatial transfer functions (STFs) which are spatial Fourier transforms of the PSFs shown in Fig. 2. It shows that the 1/e width of the response in k-space is $k_r < 1.0$ cm$^{-1}$ and $k_z < 0.6$ cm$^{-1}$.

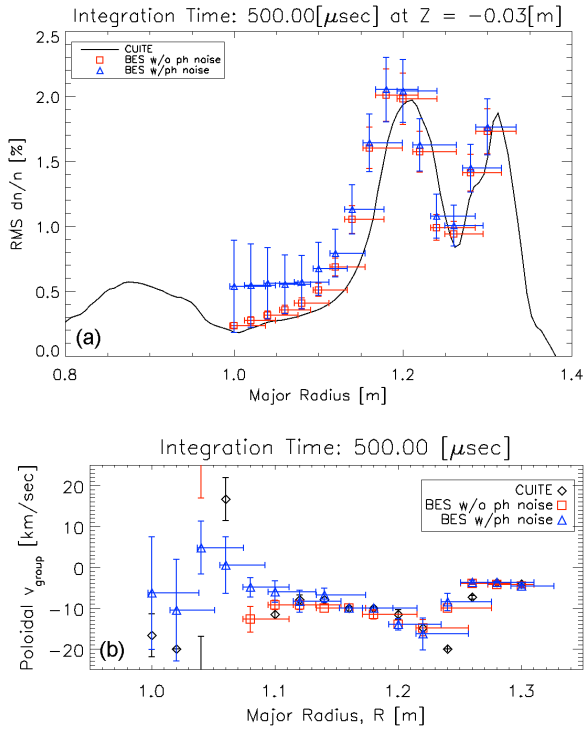

FIG. 4. (Colour online) Comparisons of (a) RMS density fluctuation level and (b) poloidal group velocity of density fluctuation between input density data from CUTIE (black) and from the synthetic BES signal (red for without photon noise and blue for with photon noise) generated from the CUTIE data on the PSF.

The PSFs generated on DIII-D can be readily used to obtain the true signal because their PSFs do not strongly vary in space.[12] However, deconvolution of the measured signal using the calculated PSFs on MAST is not trivial as the functions vary in space. This is due to the much larger variation of field line pitch on a spherical tokamak compared to that on a conventional tokamak. More sophisticated algorithms will be required to enable the deconvolution with spatially varying PSFs.[14]

Without any deconvolution of the signal, some illustrative analyses are performed and compared with the original CUTIE data. RMS density fluctuation levels and poloidal group velocity of the fluctuations are shown in Fig. 4 as a function of $R_{major}$. Poloidal group velocities are calculated using a time-delay method.[2] Because the simulation calculates the absolute number of detected photons, photon noise can be added to the synthetic BES data by generating random numbers which follow a Poisson distribution whose standard deviation is set to the square-root of the detected number of photons in an integration period of 1 μs. Fig. 4 shows the statistical results of CUTIE, and synthetic BES data without and with photon noise. Without photon noise, the synthetic BES data agree with the original CUTIE data even if deconvolutions are not performed as the correlation length of turbulence simulated by CUTIE is comparable to the spatial resolution of the BES. However, photon noise creates large deviation from the actual data, becoming worse for points towards the core. It may be seen that an RMS density fluctuation level less than 0.5 % cannot be resolved as shown in Fig. 4(a). This is because of the smaller local beam densities ($n_b$) in the core of MAST due to the beam attenuation as the beam passes the plasma. However, RMS density fluctuations below 0.5 % could be resolved in the region of higher $n_b$. The same argument can be carried to the measurement of poloidal group velocities. Note that poloidal group velocities from the simple time-delay method provide surprisingly good agreement in the edge region even if poloidal spatial resolution is degraded. This is important because time-resolved poloidal group velocities are necessary to detect GAMs and zero-mean-frequency zonal flows.[4,5] Further enhancement of signal to noise ratio on the BES data can be achieved using the correlation analysis technique if the frequency bandwidth of the noise is much broader than that of the density fluctuation.[15] This technique sets the detection limit of fluctuations about a factor of 10 below the photon noise level.

## IV. CONCLUSIONS

With the aim of proper quantitative analyses of new 2D BES data on MAST, the PSFs are generated including the appropriate physical effects. It is found that the radial spatial smearing does not depend significantly on the observation locations; whereas the poloidal spatial smearing becomes greater as the observation locations are moved towards the edge of MAST due to increasing magnetic field pitch angles. The finite half-life of $D_\alpha$ plays an equally significant role to determine the radial spatial smearing effects. The PSFs at different major radii vary which will complicate the deconvolution of the measured signal. However, even if deconvolution is not performed, the measured signal provides useful results. Effect of photon noise cannot be ignored in the core of MAST where both local beam density and plasma density fluctuations are small.

## ACKNOWLEDGEMENTS

Ghim(Kim) would like to thank Dr. Alexander Schekochihin, Dr. Martin Valovič, Dr. George McKee, and Dr. Troy Carter for valuable comments and discussions. This work was funded jointly by the UK EPSRC, by the European Communities under the contract of Association between EURATOM and CCFE and by the Leverhulme Trust International Network for Magnetized Plasma Turbulence. The views and opinions expressed herein do not necessarily reflect those of the European Commission.